\documentclass[10pt]{article}

\usepackage{amsmath}
\usepackage{amssymb}

\usepackage{graphicx,lscape,lineno}

\usepackage{cite}

\usepackage{color} 

\usepackage{setspace} 

\usepackage[labelfont=bf,labelsep=period,justification=raggedright]{caption}

\makeatletter
\renewcommand{\@biblabel}[1]{\quad#1.}
\makeatother

\date{}

\begin{document}

\begin{flushleft}
{\Large
\textbf{Quantifying the behavioural relevance of hippocampal neurogenesis}
}
\\
Stanley E. Lazic$^{1,\ast}$, 
Johannes Fuss$^{2}$, 
Peter Gass$^{3}$
\\
\bf{1} In Silico Lead Discovery, Novartis Institutes for Biomedical Research, 4002 Basel, Switzerland
\\
\bf{2} Department of Psychiatry and Psychotherapy, Central Institute of Mental Health,  University of Heidelberg, Medical Faculty Mannheim, Mannheim, Germany
\\
\bf{3} Department of Psychiatry and Psychotherapy, Central Institute of Mental Health,  University of Heidelberg, Medical Faculty Mannheim, Mannheim, Germany
\\
$\ast$ E-mail: stan.lazic@cantab.net
\end{flushleft}

\section*{Abstract}
Few studies that examine the neurogenesis--behaviour relationship formally establish covariation between neurogenesis and behaviour or rule out competing explanations. The behavioural relevance of neurogenesis might therefore be overestimated if other mechanisms account for some, or even all, of the experimental effects. A systematic review of the literature was conducted and the data reanalysed using causal mediation analysis, which can estimate the behavioural contribution of new hippocampal neurons separately from other mechanisms that might be operating. Results from eleven eligible individual studies were then combined in a meta-analysis to increase precision (representing data from 215 animals) and showed that neurogenesis made a negligible contribution to behaviour (standarised effect = 0.15; 95\% CI = -0.04 to 0.34; p = 0.128); other mechanisms accounted for the majority of experimental effects (standardised effect = 1.06; 95\% CI = 0.74 to 1.38; p = 1.7 $\times 10^{-11}$).

\section*{Introduction}

There is a consensus that new neurons in the hippocampus have a causal and biologically significant influence on cognitive and affective behaviours. However, three conditions commonly cited for establishing causation---temporal precedence, covariation, and the elimination of competing explanations---are rarely satisfied by studies that examine the neurogenesis--behaviour relationship. Temporal precedence is the only condition that is usually met because experimental interventions are applied before behavioural assessments, and changes in neurogenesis occur in between ($Experimental~intervention \rightarrow changes~in~neurogenesis \rightarrow behavioural~assessment$). Few studies quantitatively estimate the covariation between neurogenesis and behaviour; most only provide a qualitative description of a \textit{group level} association by demonstrating that the experimental group with greater neurogenesis had better behavioural performance, or vice versa. This is not the same as demonstrating that \textit{individual animals} with higher levels of neurogenesis have better behavioural performance. Group level associations cannot be used to infer individual level associations---this is known as the ecological fallacy \cite{Robinson1950}. It is even possible for associations at the group level to differ in sign from those at the individual level. The few studies that perform an individual level analysis unfortunately often do the correlation or regression through all of the data, without regard for the experimental groups. This rarely tests a hypothesis of interest and can lead to incorrect inferences \cite{Bewick2003,Hassler2003}. 

\begin{figure}[!ht]
\begin{center}
\includegraphics[scale=0.75]{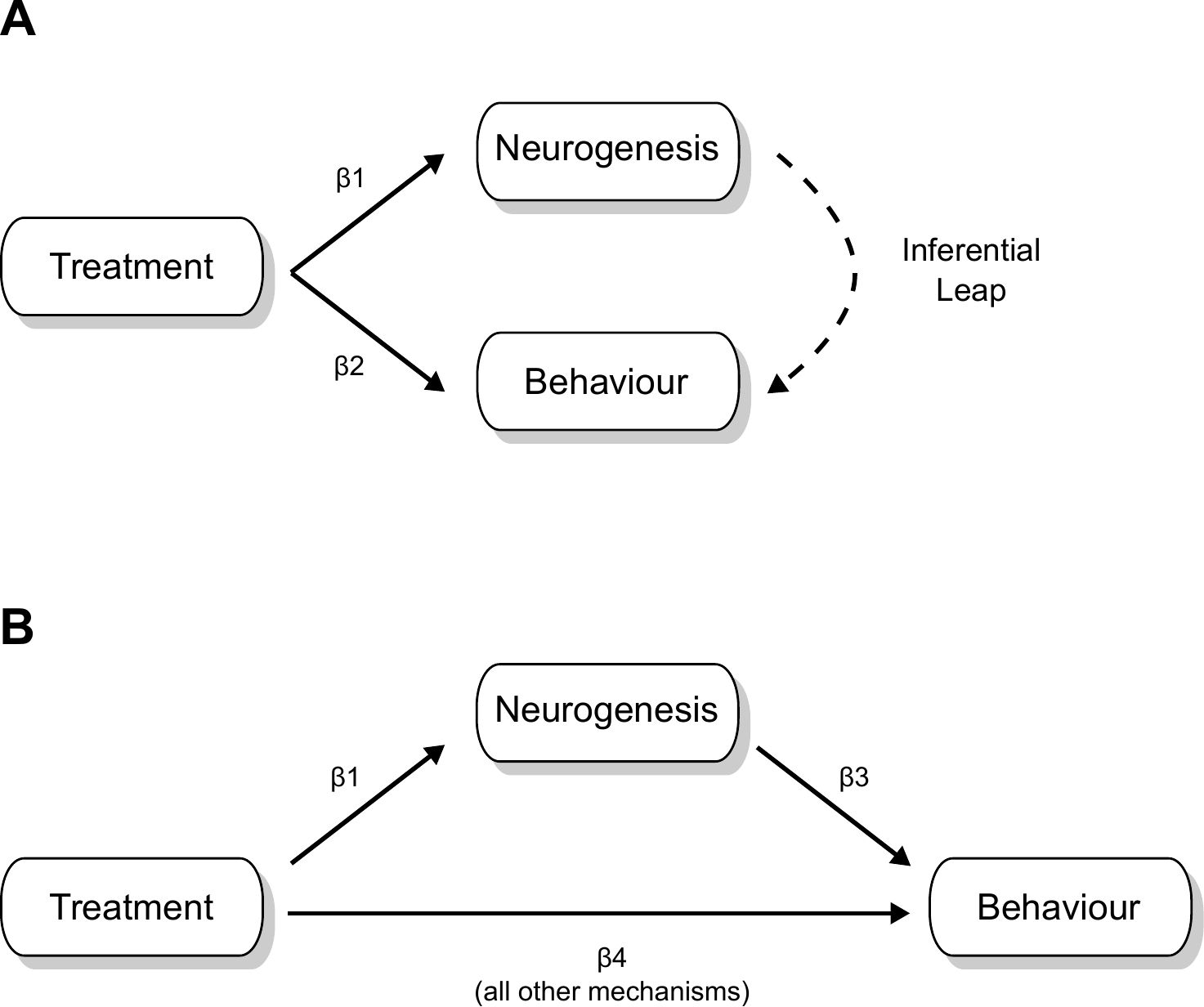}
\end{center}
\caption{{\bf Relating neurogenesis to behaviour.} A typical analysis (A) only demonstrates that a treatment or experimental intervention affects neurogenesis (p-value for $\beta_1$ is $<0.05$) and behaviour (p-value for $\beta_2$ is $<0.05$). An unjustified inferential leap is then made by concluding that changes in neurogenesis are responsible for changes in behaviour (dashed arrow), thus ignoring other potential explanations. The appropriate mediation analysis (B) estimates the effect of the treatment that is mediated by neurogenesis ($\beta_1 \times \beta_3$) and via all other mechanisms ($\beta_4$).}
\label{fig:models}
\end{figure}

There is a deeper issue in that the experimental interventions used to alter levels of neurogenesis (e.g. exercise, stress, irradiation, hormones, etc.) have numerous effects besides altering neurogenesis (see Table 1 in reference \cite{Lazic2012}), which may influence behaviour and therefore become competing explanations. Indeed, the more thorough studies measure multiple histological or physiological properties of the hippocampus and often find differences between experimental groups. Each of these outcomes becomes a competing explanation of how a particular experimental intervention affects behaviour. Few studies in the neurogenesis literature directly test whether changes in behaviour can be attributed to changes in neurogenesis and conclusions are based on an unjustified inferential leap (Fig. \ref{fig:models}A). The typical study demonstrates (1) that the experimental intervention affects neurogenesis, (2) that the intervention affects behaviour, and then inevitably concludes (3) that neurogenesis was responsible for the behavioural change (dashed line). Few recognise that no conclusion can be made about the relevance of neurogenesis with this analysis and reasoning. To see why the form of reasoning is fallacious one simply has to replace neurogenesis with any other variable: ``older mice (1) have more grey hair and (2) perform worse on the memory task, therefore we conclude (3) that hair colour is important for memory performance''. Arguing that neurogenesis is a plausible explanation (unlike hair colour) because of its neuroanatomical location, the electrophysiological properties of new neurons, etc. is assuming the point to be proved. The influence of other mechanisms is highly likely and could account for some---or even all---of the observed behavioural effects. Perfunctory caveats in the discussion noting that other mechanisms might also be involved do not adequately address the issue because if the majority of the behavioural effects are due to other mechanisms then the conclusion about neurogenesis' relevance needs to be reconsidered. Many of these competing explanations can be eliminated by using a mediation analysis, which separates the total effect of an intervention into the part due to neurogenesis and the part due to all other mechanisms (Fig. \ref{fig:models}B), and such methods are used routinely in many fields. 

Recent reviews on the neurogenesis--behaviour relationship have classified studies as providing either correlational or causal evidence, although the criteria used to make this distinction are often unspecified and differ between authors \cite{Zhao2008,DeCarolis2010,Koehl2011,Petrik2012,Vadodaria2014}. We argue that this distinction is both artificial and unhelpful for three reasons. First, both classes of studies (however defined) are of the same design: they either randomise animals to different treatment conditions or use pre-existing groups such as old and young animals. The difference between groups on neurogenesis and behaviour is then examined. The only difference between studies is the probability that the observed behavioural effects are actually related to neurogenesis and not to some other effect of the intervention. A study is not causal because neurogenesis was increased using a highly selective genetic approach and correlational because physical activity was used. Second, even the so-called causal studies still make inferences by qualitatively describing group level associations. Third, the so-called correlational studies can provide causal evidence by separately estimating the effect of neurogenesis from other effects with a mediation analysis.

Previous studies that used a mediation analysis or related methods did not show a significant effect of neurogenesis, and the estimates were close to zero \cite{Bizon2004,Lazic2010a,Castro2010,Lazic2012}. This suggests that the relevance of neurogenesis on various behavioural tasks is much more modest than commonly believed (or perhaps even non-existent) and that other mechanisms account for the observed effects. We therefore sought to obtain a better understanding of the neurogenesis--behaviour relationship by reanalysing all publicly available data.  A Bayesian mediation analysis was used to estimate neurogenesis' unique contribution and then results combined across studies to increase the precision of the estimates. Neuroscience studies often have small sample sizes and thus low statistical power \cite{Button2013} and most published studies are underpowered to detect any role that neurogenesis may have when analysed with a mediation model \cite{Fritz2007,Thoemmes2010}. However, by integrating data across studies the low power of individual experiments is irrelevant as they are combined to obtain precise estimates. The combined sample size of 215 makes this one of the largest neurogenesis analyses to date. All animal models, methods of manipulating neurogenesis, behavioural outcomes, and study designs were eligible for inclusion.  The quality of the literature was also assessed by quantifying the number of studies that used randomisation, blinding, took litter-effects into account, whether data points were dropped without mention or justification, and whether there was selective reporting of neurogenesis--behaviour associations, as these are known to introduce bias in the reported results.

\section*{Results and Discussion}

\subsection*{Neurogenesis' contribution to behaviour is minimal}
\begin{landscape}
\begin{figure}[ht]
\begin{center}
\includegraphics[scale=0.67]{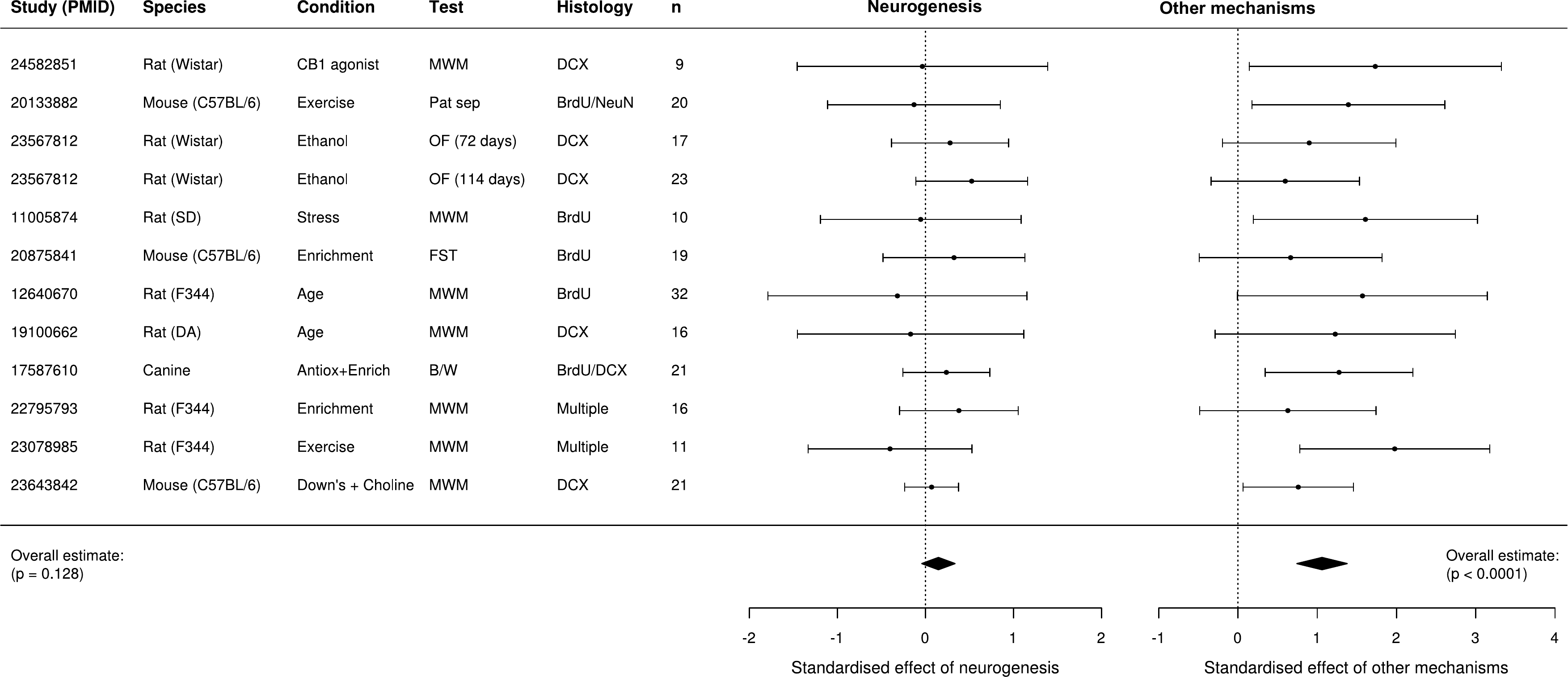}
\end{center}
\caption{{\bf Study specific and combined estimates from the causal mediation analysis.} The effect of neurogenesis was smaller than the effect of other mechanisms in all studies. The overall effect of neurogenesis was small, non-significant (p = 0.128), and unlikely to account for a large proportion of the observed behavioural effects (upper 95\% CI = 0.34). PMID = PubMed ID; DA = Dark agouti; SD = Sprague-Dawley; MWM = Morris water maze; OF = open field; FST = forced swim test; B/W = black/white discrimination task. Error bars are 95\% credible intervals.}
\label{fig:meta_results}
\end{figure}
\end{landscape}

Figure \ref{fig:meta_results} displays the individual and combined effect estimates from neurogenesis and other mechanisms, and it is clear that the effect of other mechanisms was larger than that of neurogenesis in all cases. The black diamonds are the combined estimates across the eleven studies. Overall, neurogenesis made a small---and some might argue negligible---12\% contribution to behaviour that was not significant (standardised effect = 0.15; 95\% CI = -0.04 to 0.34; p = 0.128) and non-neurogenesis mechanisms played a much greater role (standardised effect = 1.06; 95\% CI = 0.74 to 1.38; p = 1.7 $\times 10^{-11}$). While not part of the literature search, one additional study was available to us \cite{Fuss2010}, which had a greater effect of neurogenesis compared to other mechanisms. Including this study in the analysis increases the overall effect of neurogenesis from 0.15 to 0.18 (95\% CI = -0.01 to 0.36; p=0.062). Assuming that neurogenesis has some effect, including additional studies will eventually produce a statistically significant result and the lower confidence limit will exclude zero. Of greater interest however is the upper confidence limit, which places an upper bound on the magnitude of neurogenesis' contribution. It is clear in Figure \ref{fig:meta_results} that large or even moderate effects of neurogenesis are not supported by the data, properly analysed. The conclusions of these studies therefore needs to be revisited, and it should be noted that the degree to which the authors believed that their study provided support for a neurogenesis--behaviour relationship varied greatly.

An advantage of a meta-analysis is that the consistency of effects across studies can be estimated \cite{Borenstein2009}. It is clear that the effect of neurogenesis is consistently small across a variety of different species, behavioural outcomes, and methods of manipulating neurogenesis (Fig. \ref{fig:meta_results}), which is confirmed with a test for between-study heterogeneity (empirical Bayes estimate: $\tau^2$ = 0; SE = 0.048; Q$_{(11)}$ = 4.98, p = 0.932). If the estimated contribution of neurogenesis differed greatly between studies then follow-up analyses could attempt to attribute this variation to properties of the studies. For example, are some behavioural tests more sensitive to changes in neurogenesis than others, or is neurogenesis more important in certain species or strains? Despite differences in the characteristics of the experiments, they all consistently showed little or no effect of neurogenesis. This means that studies not included in the meta-analysis would be unlikely to show a dramatic effect of neurogenesis. Even though only a small proportion of published studies could be included in the meta-analysis the low between-study variability strengthens the generalisability of the results.

\subsection*{Quality of the literature}
Only 36\% (4/11) of studies reported using blinding, 36\% (4/11) used randomisation, and 18\% (2/11) commented on how the design of the study accounted for potential litter effects. To our knowledge, no studies have examined or reported on litter-to-litter variation in neurogenesis, but litter effects are present for many outcomes (including behaviour; see reference \cite{Lazic2013} and references therein) and are a source of both bias and noise. Given the large number of factors that affect neurogenesis, it would be surprising if it were stable across litters. In addition, 45\% (5/11) of the studies appeared to have incomplete data in that the number of data points in the figures was less than the number of animals mentioned in the methods section, or the reported degrees of freedom did not match the indicated sample size and method of analysis \cite{Lazic2010}. Finally, 64\% (7/11) of studies selectively reported results in that only some associations between behavioural outcomes and neurogenesis were reported. Not surprisingly, all reported associations from these studies were significant. These results are similar to other preclinical studies using animal models  \cite{Bebarta2003,Macleod2004,Crossley2008,Philip2009,Kilkenny2009,Sena2010,Vesterinen2010,Rooke2011,Moja2014} and are known to inflate effect sizes (reviewed by Dirnagl \cite{Dirnagl2006}).

\subsection*{Is neurogenesis involved with forgetting?}

\begin{figure}[!ht]
\begin{center}
\includegraphics[scale=0.65]{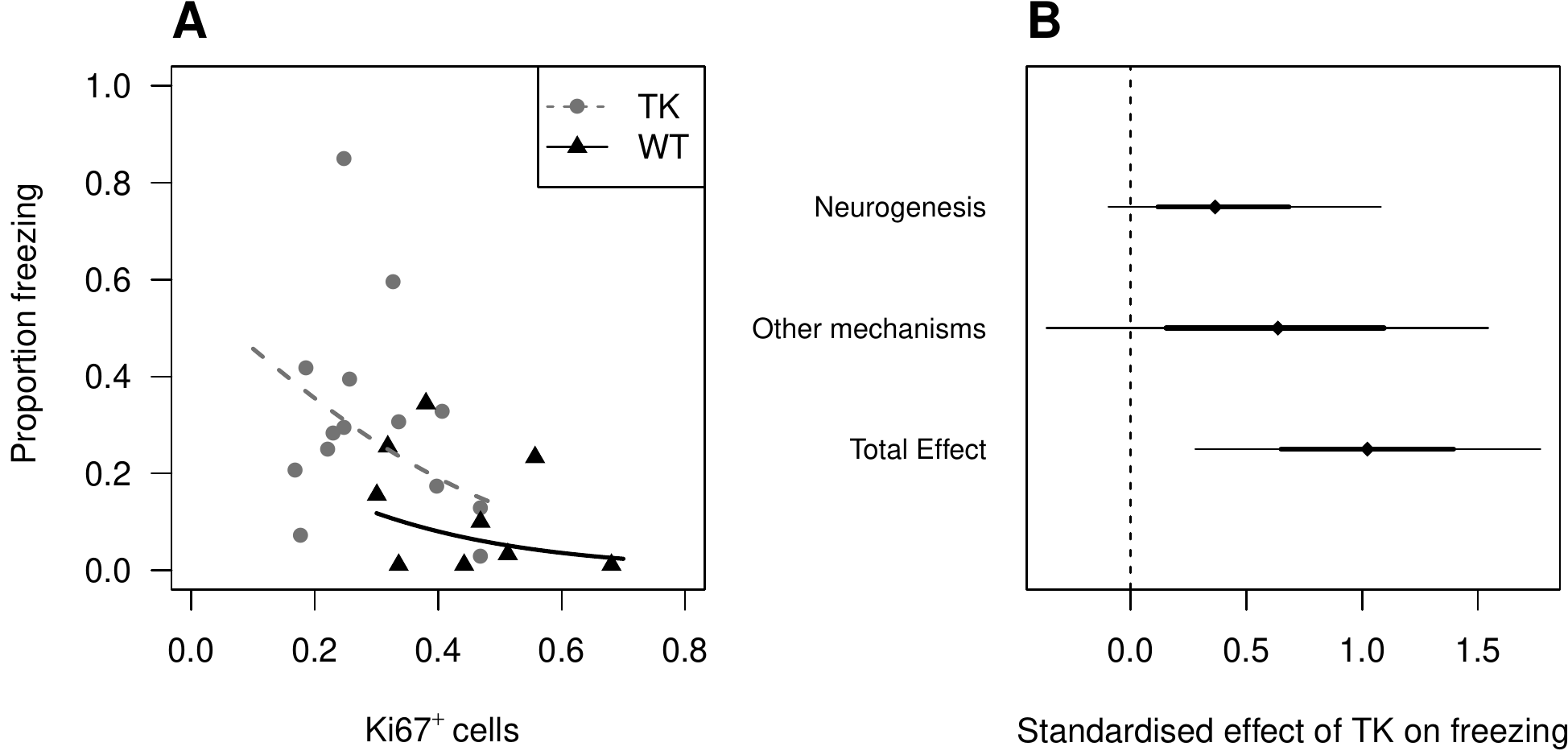}
\end{center}
\caption{{\bf Association between neurogenesis and forgetting.} (A) Raw data extracted from Figure 4F in Akers et al. \cite{Akers2014} and estimated regression lines. While in the predicted direction (negative association), the relationship is not significant within groups (p = 0.12). (B) The causal mediation analysis divides the total effect of the TK transgene into the part that can be attributed to neurogenesis and to all other mechanisms. The effect of neurogenesis accounted for less than half (40\%) of the total effect, although there is a great deal of uncertainty in this estimate. Error bars are 95\% (thin) and 50\% (thick) credible intervals.}
\label{fig:freezing}
\end{figure}

A recent paper by Akers et al. published after the previous analysis was completed suggests that high levels of neurogenesis might be associated with increased forgetting, particularly when neurogenesis is increased after learning has taken place \cite{Akers2014}. This is a different proposed behavioural function for neurogenesis compared to previous studies.  Only data for one experiment could be extracted (Figure 4F in the paper), which used a genetic knock-down of neurogenesis. Wild-type (WT) mice were compared with transgenic mice expressing thymidine kinase (TK) with a nestin promoter. Administering ganciclovir to TK mice causes apoptosis of nestin expressing cells, thus reducing the number of new neurons. The primary outcome in this paper was the length of time animals were immobile or frozen when placed into an environment in which they had previously received an electrical shock, and was expressed as the percent of total time in the test environment. Since the outcome is bounded between 0 and 100\%, the values were transformed to the logit scale ($log\left( p/(1-p) \right)$ after dividing by 100 to convert percents into proportions. This allows a normal linear model to be used for the reanalysis; otherwise predictions and confidence intervals would include impossible values (outside of the [0,1] interval). The data were then back transformed onto the proportion scale for plotting (Fig. \ref{fig:freezing}A).

If the experimental groups are ignored, there is an association between the number of Ki67$^+$ cells and the proportion of time spent freezing, which would appear to provide evidence for a causal role for neurogenesis. However, the direction and strength of this association is irrelevant for addressing the question of whether neurogenesis is important for behaviour---a point which has not been appreciated by the majority of the neurogenesis community. In order to explain why, an unrelated example will be used so that prior opinions do not influence the interpretation. Suppose we have the hypothesis that taller people earn higher incomes, and specifically that greater height \textit{causes} greater income. To simplify the analysis we could examine males only and one profession only, say that of barbers. Collecting data on the height and income of a number of barbers would then allow us to address the association between these variables, which would be a partial test of the hypothesis by establishing whether covariation exists. Suppose that the data were collected from two countries: the Netherlands (tall and high income by global standards) and individuals from a pygmy population in the Congo (short \cite{Bozzola2009} and low income). A graph of height ($x$-axis) versus income ($y$-axis) would show a cluster of points in the top right (Dutch population) and in the bottom left (African population) and a correlation or regression through all of the data (ignoring the country that individuals are from) would show an impressive positive association and a small p-value. This however is irrelevant for our hypothesis: the Dutch barbers earn higher incomes not because they are so tall but because they live in a modern industrialised country. If height was causally involved in determining income, then we would expect this association to hold \textit{within} the two populations---the tallest Dutchman should earn more than the shortest and the tallest Congolese should earn more than the shortest. It is the association within groups that matters (which can be tested in a single analysis with an analysis of covariance for example and does not require a separate analysis for each population \cite{Lazic2012}). Similarly, it is the association between neurogenesis and behaviour within the WT group and within the TK group that tests the causal hypothesis, not the overall association. This generalises to any experimental or grouping variable.

In the Akers data set, the association within the experimental groups is still negative but not significant (p = 0.12), which might be due to lower power. A mediation analysis was also performed on the standarised variables and the effect of neurogenesis was 0.399 (95\% CI = -0.10 to 1.08), which is 40\% of the total effect (Fig. \ref{fig:freezing}B). The standarised effect attributed to all other mechanisms was 0.623 (95\% CI = -0.36 to 1.54). It is clear from the 95\% CI that there is a great deal of uncertainty in the estimates and that a larger sample size is required, but it is clear that there is no overwhelming effect of neurogenesis. It might be surprising that genetic knock-down of neurogenesis still has such a large effect attributed to other mechanisms, as this method of manipulating neurogenesis is less likely to have additional off-target effects. Some of this might be attributed to biological or technical variables that are affecting the results such as litter effects, or the sex of the person handling the animals \cite{Sorge2014}.

There are several points to consider. First, these results show that the behavioural effects of experimental interventions are driven mainly by neurogenesis-independent mechanisms. The studies included in the combined analysis are representative of the literature and include several strains of rats and mice, measures of both cognitive and affective behaviour, and many of the standard methods of manipulating neurogenesis (e.g. exercise, environmental enrichment, stress, and age). It is therefore likely that the relevance of neurogenesis in general has been greatly overestimated, bringing into question the usefulness of increasing neurogenesis for therapeutic purposes.

Second, these results should not be at all surprising because it is known that the experimental interventions typically used to alter levels of neurogenesis have many other effects that influence behaviour \cite{Lazic2012}, and thus they are potential alternative explanations that need to be ruled out to make a convincing case for neurogenesis. For example, Lipp and colleagues showed a high within group correlation between the extent of intra- and infrapyramidal mossy fibre projections from the dentate to the CA3 and performance on a hippocampal dependent behavioural task \cite{Lipp1988}. This was evident in the control group as well, suggesting that natural variation in mossy fibre projections are a good predictor of behaviour, and any experimental intervention that affects this will possibly influence behaviour, regardless of the intervention's effect on neurogenesis. Behaviour is a ``final common pathway'' of countless factors and biological processes, making it \textit{a priori} unlikely that any single factor will dominate. 

Third, one could argue that the experiments examined here are too simple to establish causation, and more complex experiments that both increase (e.g. exercise) and decrease (e.g. irradiation) neurogenesis are required, thus ruling out other explanations by experimental means. It is still possible however to be misled with these ``$2\times 2$'' designs when using standard statistical methods \cite{Lazic2012}. The problem lies in how inferences are made rather than the experimental design. Many studies attempt to address the issue of off-target effects by using multiple methods to alter levels of neurogenesis. While this provides a qualitative way to increase confidence in the relevance of neurogenesis, a series of biased estimates still does not adequately address the issue.

The fourth point relates to limitations of the analysis. The mediation model does not provide proof of neurogenesis' role in behaviour (even if the effect was significant), as there might be another variable which is mediating the effect of the intervention and neurogenesis just happens to be correlated with it \cite{Fiedler2011}. This analysis can provide a necessary but not sufficient condition for demonstrating a causal role for neurogenesis. The three-variable model used here does not capture the full biological complexity, and future studies should include other variables in more detailed and realistic models. This would provide a better understanding of how multiple mechanisms interact and their relative importance for behaviour. For example, the effect of neurogenesis might be underestimated if there are compensating mechanisms that counteract the effect of increased or decreased neurogenesis. An assumption of the mediation analysis is that the behavioural training and testing has no---or at least a negligible---effect on neurogenesis. This will likely not be a problem however unless the effect of training and testing on neurogenesis differs between experimental groups. A final point is that the mediation models used here assume that all groups have at least some animals with neurogenesis. The models would not work well if all animals in the experimental group had a cell count of zero for example, as there needs to be some variation in neurogenesis to account for variation in behaviour.

Fifth, the results of the combined analysis are only as good as the studies that go into it. Remarkably few studies reported the covariation between neurogenesis and behaviour (which explains why so few studies were included despite a large number of published papers), but it is an obvious result to mention since it directly addresses the research hypothesis. The results are likely biased in favour of neurogenesis if studies only displayed the association between neurogenesis and behaviour if it was statistically significant and in the expected direction; and even studies that reported these results did so selectively in that not all neurogenesis--behaviour associations were mentioned. Publication bias is therefore a concern and is known to be a problem in the preclinical biomedical literature \cite{Sena2010,Tsilidis2013}. The neurogenesis literature also suffers from the problem of multiplicity: multiple behavioural tests are often conducted, with multiple outcomes for each behavioural test, multiple markers of proliferation or new neurons are used (e.g. Ki67, BrdU, BrdU/NeuN, DCX, BrdU/DCX, etc.), and multiple divisions of the hippocampus are possible (the whole dentate gyrus, dorsal versus ventral, or infrapyramidal versus suprapyramidal blade, etc.). There may be good reasons for making such anatomical distinctions and the point is that there are many neurogenesis--behaviour correlations that can be conducted per study. Such multiple testing increases the chance of false positives and is rarely taken into account. To this one can add variations in analysis such as with and without assuming equal variances, log transformations, non-parametric methods, the removal of inconvenient data points, etc. This flexibility in choice of analysis has been referred to as ``researcher degrees of freedom'' and can also increase the number of false positives \cite{Simmons2011}. It is also rare for studies to define primary histological and behavioural outcomes before conducting the study, and thus any significant result is taken as evidence that neurogenesis is relevant for behaviour. After some fifteen years of neurogenesis research, surely we are in a position to define primary outcomes before conducting the experiment---in other words, to make a specific prediction about the expected relationships \cite{Wagenmakers2012a,Kimmelman2014}. The above factors are recognised as important for the reproducibility of research \cite{Landis2012}, as is the application of appropriate statistical methods to make valid conclusions, rather than inferential leaps unsupported by the data.

Finally, while this article has been mostly critical, it is worth reflecting on some positive aspects. Compared with other fields where causal explanations are sought, experimental biologists have a number of advantages, including the use of randomisation, a high degree of experimental control, the ability to create large effects, access to homogeneous sample material, ease of reproducing the treatment across subjects (i.e. all animals ``comply'' with the treatment), and dealing with directly observable phenomena such as cell counts and behaviour. These advantages---coupled with statistical models that directly estimate the hypothesis of interest---provide a powerful set of methods for understanding brain--behaviour relationships.

\section*{Materials and Methods}

To find relevant data a PubMed search using the phrase ``neurogenesis AND (dentate OR hippocamp*) AND (behavior OR behaviour) AND English [LA] NOT review [PT]'' was conducted and returned 990 papers, and included publications up to 4 April 2014. The flow diagram can be found in the Supporting Information (Text S1). Not all 990 papers were relevant as some were review articles while others were \textit{in vitro} studies. The papers were examined for \textit{in vivo} studies that either assigned animals to different treatment conditions or used naturally occurring groups such as old versus young. Otherwise, all animal models, methods of manipulating neurogenesis, behavioural outcomes, and study designs were eligible for inclusion. However, only data presented as scatterplots (neurogenesis versus behaviour) could be extracted and therefore used. None of the studies provided the raw data as supplementary material, otherwise they would have been included. In addition, it was necessary for each animal to be distinguished by experimental group; for example, with different colours or shapes for the plotting symbols, or if the groups were plotted in separate panels or figures. Only fifteen studies remained after applying the above criteria and the data were accurately extracted from the figures using g3data software (www.frantz.fi/software/g3data.php). Two studies were subsequently excluded as over-plotting of points made it impossible to extract the data in one, and the control group was omitted from the scatterplot in the second. Two additional papers had a strong group $\times$ neurogenesis interaction on the behavioural outcome. These were not included in the combined analysis but are discussed in the Supporting Information (Text S1). There were therefore eleven studies in the final analysis with 215 animals.

The mediation models were implemented as Bayesian graphical models and were fit in R (version 3.0.3) using the R2jags package (version 0.04-01) and JAGS (Just Another Gibbs Sampler; version 3.4.0). Measures of neurogenesis and behaviour were standardised to have a mean of zero and standard deviation of one, which allowed the effects to be pooled across studies. Note that this only puts the results of the studies on a common scale, but does not otherwise change them. The results are therefore interpreted in standard deviation units; for example, an estimated effect of 0.2 for neurogenesis means that the experimental intervention via its effect on neurogenesis increases behaviour by 0.2 standard deviations relative to the control group. A similar interpretation applies for the neurogenesis-independent effect. For some behavioural outcomes a lower value indicates better performance and therefore the estimated effect of the intervention would be negative. In such cases the sign of the effect was reversed in the combined analysis so that positive estimates are always associated with better performance. The Markov chain Monte Carlo (MCMC) sampling used three chains of 1,000,000 iterations each, a burn-in period of 5000 iterations, and every tenth value was saved. The three chains were well mixed (Gelman-Rubin statistic $\leq$1.01 for all parameters). The main parameters of interest were the effect of the treatment or group on behaviour via neurogenesis-independent mechanisms ($\beta_4$ in Fig. 1B) and the effect via neurogenesis (which is equal to the product of coefficients $\beta_1 \times \beta_3$). Non-informative priors were used and the results were not sensitive to the form of the prior. Further details of this modelling approach applied to neurogenesis data can be found in Lazic \cite{Lazic2012}, and a general introduction to mediation analysis can be found in references \cite{Baron1986,Shrout2002,MacKinnon2007,MacKinnon2009,Hayes2009,Yuan2009,Hayes2013}. Causal models are discussed in detail by Pearl \cite{Pearl2009} and Bayesian graphical models by Koller and Friedman \cite{Koller2009}.

After the estimates were obtained from each of the eleven studies they were combined with a random effects meta-analysis using the metafor R package (version 1.8-0; \cite{Viechtbauer2010}). The studies often contained multiple measures of neurogenesis, behavioural tests, and outcomes for a particular behavioural test (e.g. latency to immobility and total immobility time on the forced swim test). To avoid ``double counting'' only the most precise estimate from each study was used in the combined analysis, but all estimates are available in the Supporting Information (File S1). The data extracted from the published manuscripts are also provided in the Supporting Information (File S1) along with the R code. A protocol for this analysis has not been published.

The quality of the eleven studies that went into the combined analysis was assessed using five metrics that are known to introduce bias \cite{Hooijmans2014}. First, the number of studies mentioning random allocation of animals to different treatment groups (if applicable) was determined. Second, the number of studies indicating that the experimenter was blind when quantifying neurogenesis or when assessing behavioural outcomes was determined. The criteria used was generous in that if blinding was mentioned for only one of the above, the study was counted as having been blinded. Third, the number of studies that discussed how potential litter effects were handled was determined. Litter effects occur because littermates tend to be more alike on a variety of outcomes compared with animals from different litters. In other words, the variation between litters is greater than the variation within litters \cite{Lazic2013}. If experiments are not designed with litter effects in mind, both the false positive and false negative rates can be dramatically increased. Fourth, a check on data consistency was performed by examining whether data points were removed without comment. This can be determined by inconsistencies in numbers of data points across relevant figures, a discrepancy between the number of animals reported in the methods compared with those shown in the figures, or when the reported degrees of freedom do not match the indicated sample size and method of analysis \cite{Lazic2010}. Finally, selective reporting was defined as the reporting of only a subset of all possible neurogenesis--behaviour associations. For example, if three behavioural outcomes are described, but the result of only one association with neurogenesis is reported. Or, if multiple measures of neurogenesis are described (e.g. Ki67 and DCX) but the result of only one association with behaviour is reported. The quality metrics were not used in any way in the meta-analysis.

\section*{Acknowledgements}

We thank the anonymous reviewers for constructive comments.

\bibliography{neurogenbib}

\clearpage
\section*{Figure Legends}

\section*{Supporting Information Legends}


\begin{description}
\item {\bf Text S1. Supplementary information.} Contains the meta-analysis flow diagram and a discussion of two studies with strong group $\times$ neurogenesis interaction effects.
\item {\bf File S1. Data files and R code.}
\end{description}

\end{document}